%% file: main.tex
\PassOptionsToPackage{hypertexnames=false}{hyperref} 
\documentclass[copyright,creativecommons]{eptcs}
\providecommand{\event}{SCSS 2021} 
\usepackage{underscore}           
\usepackage[T1]{fontenc}
\usepackage[utf8]{inputenc}

\usepackage{graphicx}
\usepackage{color}
\usepackage{csquotes}
\usepackage{newunicodechar}
\usepackage{fancyvrb}

\usepackage{mathbbol}
\usepackage{amsmath}
\usepackage{stix}
\usepackage{float}

\input{commands}

\input{unicode}

\allowdisplaybreaks

\begin{document}
\title{First-Order Logic in Finite Domains:
Where Semantic Evaluation Competes with SMT Solving}

\def\titlerunning{First-Order Logic in Finite Domains}

\author{Wolfgang Schreiner\thanks{Supported by 
the JKU Linz LIT Project LOGTECHEDU and 
by the Aktion Österreich-Slowakei Project 2019-10-15-003.}
\institute{Research Institute for Symbolic Computation (RISC) \\ Johannes
Kepler University Linz, Austria} 
\email{Wolfgang.Schreiner@risc.jku.at}
\and 
Franz-Xaver Reichl\thanks{Supported by the Austrian Science Fund (FWF) 
under grant W1255.}
\institute{Algorithms and Complexity Group \\ TU Wien, Austria }
\email{freichl@ac.tuwien.ac.at}
}

\def\authorrunning{W. Schreiner and F.-X. Reichl}
\def\sectionautorefname{Section}  

\maketitle            

\begin{abstract}

In this paper, we compare two alternative mechanisms for deciding the
validity of first-order formulas over finite domains supported by the
mathematical model checker RISCAL: first, the original approach of
\enquote{semantic evaluation} (based on an implementation of the
denotational semantics of the RISCAL language) and, second, the later
approach of SMT solving (based on satisfiability preserving translations of
RISCAL formulas to SMT-LIB formulas as inputs for SMT solvers). After a
short presentation of the two approaches and a discussion of their
fundamental pros and cons, we quantitatively evaluate them, both by a set of
artificial benchmarks and by a set of benchmarks taken from real-life
applications of RISCAL; for this, we apply the state-of-the-art SMT solvers
Boolector, CVC4, Yices, and Z3. Our benchmarks demonstrate that (while SMT
solving generally vastly outperforms semantic evaluation), the various SMT
solvers exhibit great performance differences. More important, we identify
classes of formulas where semantic evaluation is able to compete with (or
even outperform) satisfiability solving, outlining some room for
improvements in the translation of RISCAL formulas to SMT-LIB formulas as
well as in the current SMT technology.

\end{abstract}

\input{core}

\bibliographystyle{eptcs}
\bibliography{main}

\input{appendix}

\end{document}

%% file: commands.tex
\newcommand{\symv}[1]{\textit{#1}}
\newcommand{\symc}[1]{\textsf{#1}}

\newcommand{\symdeff}{\ensuremath{\coloneq}}

\newcommand{\symnat}{\ensuremath{\mathbb{N}}}
\newcommand{\symint}{\ensuremath{\mathbb{Z}}}

\newcommand{\symnot}{\ensuremath{\neg}}
\newcommand{\symand}{\ensuremath{\wedge}}
\newcommand{\symor}{\ensuremath{\vee}}
\newcommand{\symimp}{\ensuremath{\Rightarrow}}

\newcommand{\symequ}{\ensuremath{\Leftrightarrow}}
\newcommand{\symforall}[1]{\ensuremath{\forall{#1}.\ }}
\newcommand{\symexists}[1]{\ensuremath{\exists{#1}.\ }}

\newcommand{\symemptyset}{\ensuremath{\emptyset}}
\newcommand{\symunion}{\ensuremath{\cup}}
\newcommand{\symintersect}{\ensuremath{\cap}}
\newcommand{\symsubset}{\ensuremath{\subset}}
\newcommand{\symsubseteq}{\ensuremath{\subseteq}}

\newcommand{\symchoose}[1]{\symc{choose}\ {#1}.\ }

\newcommand{\symtype}[2]{\ensuremath{{#1}{:}{#2}}}

{\endgraf\csname shaded\endcsname\quote\verbatim}%
{\endverbatim\endquote\csname endshaded\endcsname\noindent}
  
\makeatletter
\newcommand{\verbatimfont}[1]{\def\verbatim@font{#1}}%
\makeatother

\newcommand{\symseml}{\ensuremath{\Lbrack}}
\newcommand{\symsemr}{\ensuremath{\Rbrack}}
\newcommand{\sem}[1]{\ensuremath{\symseml\,{#1}\,\symsemr}}

\DefineVerbatimEnvironment{specification}{Verbatim}
{commandchars=\\\{\},formatcom=\sffamily\upshape,xleftmargin=1.5em}

\newcommand{\progtext}[1]{\texttt{\textup{#1}}}
\DefineVerbatimEnvironment{progblock}{Verbatim}
{commandchars=\\\{\}\(\),codes={\catcode`$=3\catcode`^=7},formatcom=\ttfamily\upshape\small,xleftmargin=1.5em}



%% file: unicode.tex
\newunicodechar{∀}{\ensuremath{\forall}}
\newunicodechar{∃}{\ensuremath{\exists}}
\newunicodechar{∧}{\ensuremath{\symand}}
\newunicodechar{∨}{\ensuremath{\symor}}
\newunicodechar{⟶}{\ensuremath{\longrightarrow}}
\newunicodechar{⋀}{\ensuremath{\bigwedge}}
\newunicodechar{⟹}{\ensuremath{\Longrightarrow}}
\newunicodechar{⇒}{\ensuremath{\Rightarrow}}
\newunicodechar{⇔}{\ensuremath{\symequ}}
\newunicodechar{¬}{\ensuremath{\neg}}
\newunicodechar{⟷}{\ensuremath{\longleftrightarrow}}
\newunicodechar{≠}{\ensuremath{\neq}}
\newunicodechar{λ}{\ensuremath{\lambda}}
\newunicodechar{∈}{\ensuremath{\in}}
\newunicodechar{∉}{\ensuremath{\not\in}}
\newunicodechar{≤}{\ensuremath{\leq}}
\newunicodechar{≥}{\ensuremath{\geq}}
\newunicodechar{∩}{\ensuremath{\symintersect}}
\newunicodechar{∪}{\ensuremath{\symunion}}
\newunicodechar{⋃}{\ensuremath{\bigcup}}
\newunicodechar{⋂}{\ensuremath{\bigcap}}
\newunicodechar{⊆}{\ensuremath{\symsubseteq}}
\newunicodechar{⊂}{\ensuremath{\symsubset}}
\newunicodechar{⊈}{\ensuremath{\not\subseteq}}
\newunicodechar{⟦}{\ensuremath{\symseml}}
\newunicodechar{⟧}{\ensuremath{\symsemr}}
\newunicodechar{≔}{\ensuremath{\symdeff}}
\newunicodechar{ℕ}{\ensuremath{\symnat}}
\newunicodechar{ℤ}{\ensuremath{\symint}}
\newunicodechar{∅}{\ensuremath{\symemptyset}}
\newunicodechar{⊤}{\ensuremath{\top}}
\newunicodechar{⊥}{\ensuremath{\bot}}
\newunicodechar{⋅}{\ensuremath{\cdot}}
\newunicodechar{〈}{\ensuremath{\langle}}
\newunicodechar{⟨}{\ensuremath{\langle}}
\newunicodechar{〈}{\ensuremath{\langle}}
\newunicodechar{〉}{\ensuremath{\rangle}}
\newunicodechar{⟩}{\ensuremath{\rangle}}
\newunicodechar{〉}{\ensuremath{\rangle}}
\newunicodechar{∑}{\ensuremath{\sum}}
\newunicodechar{Σ}{\ensuremath{\Sigma}}
\newunicodechar{∏}{\ensuremath{\prod}}
\newunicodechar{Π}{\ensuremath{\Pi}}
\newunicodechar{×}{\ensuremath{\times}}

%% file: core.tex
\section{Introduction}

The aim of the RISCAL system~\cite{RISCAL} is to support the analysis of
theories and algorithms over discrete domains, as they arise in computer
science, discrete mathematics, logic, and algebra. For this purpose, RISCAL
provides an expressive specification language based on a strongly typed
variant of first-order logic in which theories can be formulated and
algorithms can be specified; nevertheless the validity of all formulas and
the correctness of all algorithms is decidable. This is because all RISCAL
types have finite sizes which are configurable by model parameters.
Therefore, a RISCAL model actually represents an infinite set of finite
models; before verifying the validity of a theorem over the infinite model
set by a deductive proof in some theorem proving environment, we can check
its validity over selected finite instances of the set by model checking in
RISCAL. The system has been mainly developed for educational purposes
\cite{Schreiner2018b} but it has also been applied in
research~\cite{Schreiner2020b}.

The basic mechanism of RISCAL for deciding the validity of formulas and the
correctness of algorithms is \enquote{semantic evaluation}, which is based
on a constructive implementation of the denotational semantics of all kinds
of syntactic phrases allowed by the language. However, since 2020, the
system also provides an alternative (and potentially much more efficient)
decision mechanism based on SMT (satisfiability modulo theories) solving,
implemented by the second author~\cite{Reichl2020,Schreiner2020c}. In this
approach, the decision of a RISCAL formula is performed via a translation to
a formula in the SMT-LIB language~\cite{SMTLIB} and the application of some
external SMT solver (currently, the SMT solvers Boolector, CVC4, Yices, and
Z3 are supported). Indeed, this has achieved great performance
improvements~\cite{Reichl2020} and allowed to constructively work with
theories that were out of reach of semantic evaluation.

Actually, SMT solvers have served for a long time as backends of various
program verification tools, for real languages such as
Java~\cite{Ahrendt2016} as well as for algorithmic languages such as
Dafny~\cite{Leino2016} where the verification backend
Boogie~\cite{Barnett2005b} generates SMT-LIB conditions that are discharged
by the SMT solver Z3. Furthermore, SAT/SMT solvers are applied for the
analysis of system modeling languages such as Alloy~, Event-B, and VDM. Last
but not least, they are employed as backends for interactive provers, e.g.,
in Isabelle's \enquote{sledgehammer} component~\cite{Blanchette2011} and in
Coq's SMTCoq plugin~\cite{Ekici2017}.

As for the translation of higher-level specification languages into the
languages of satisfiability solvers, \cite{Jackson2000} describes the
techniques used in the Alloy Analyzer to transform formulas from first-order
relational logic; \cite{Torlak2007} discusses improvements implemented in
the SAT-based relational model finder Kodkod. On top of Kodkod, the
counterexample generator Nitpick~\cite{Blanchette2010} generates finite
countermodels of Isabelle formulas by a translation to relational logic. In
\cite{Ghazi2015}, it is briefly sketched how Alloy constraints have been
(manually) translated into the language of the SMT solver Yices, which shows
drastic speedups when tautologies are decided. \cite{Lin2017} sketches the
encoding of VDM proof obligations as SMT problems proved with the SMT solver
Z3. In somewhat more detail, \cite{Deharbe2012} discusses the implementation
of a SMT plugin for the Event-B platform Rodin and experimentally compares
this plugin with those for other provers.

However, as beneficial SMT solving in general is, in our own work of
deciding RISCAL formulas we also have regularly encountered cases where the
performance of the SMT-based decision is comparatively poor, sometimes even
beaten by semantic evaluation. While some potential reasons have already
been outlined in~\cite{Reichl2020}, a more systematic analysis and
evaluation has been lacking so far. Also the work reported in the scientific
literature is a bit unsatisfactory in this respect: while the relative
merits of SMT solvers are regularly evaluated in the SMT-COMP competition
series~\cite{SMT-COMP}, it is harder to find comparisons with alternative
decision mechanisms such as the one with the Vampire prover presented
in~\cite{Reger2017}.

In this paper, we provide a detailed comparison of the built-in decision
mechanism of RISCAL by semantic evaluation with the corresponding decisions
by SMT solving. In particular, we identify classes of situations where the
performance of SMT-based decisions is relatively low, i.e., where indeed
\enquote{semantic evaluation competes with SMT solving}, as a starting point
for potential improvements in the SMT-LIB translation of RISCAL and
in SMT technology in general. While our insights are clearly limited to the
particular strategy applied for translating RISCAL formulas to SMT-LIB, our
work shows that SMT is not a panacea in all kinds of reasoning problems but
has to be applied with some caveats.

Closest to our work is \cite{Merz2016}, where the untyped first-order logic
of Lamport's language TLA$^+$ is translated to SMT-LIB conditions that are
discharged by the SMT solvers CVC4 and Z3; however the results have not been
experimentally compared with the built-in TLC model checker. As another
difference, the TLA$^+$ translation heavily relies on a nonconstructive
encoding of non-integer values by uninterpreted sorts and functions with
corresponding background axioms. In contrast to this, the RISCAL translation
generates formulas over bit vectors with uninterpreted sort and function
symbols, which minimizes the use of uninterpreted functions by a
constructive encoding of all types as bit vectors.

The remainder of this paper is organized as follows: In
\autoref{sect:decide}, we outline the decision mechanisms applied in RISCAL.
In \autoref{sect:artificial}, we present the artificial benchmarks which we
use to compare both mechanisms. In \autoref{sect:real}, we extend these
investigations to a selected set of benchmarks taken from real-life
applications of RISCAL. In \autoref{sect:concl}, we present our conclusions
derived from these investigations and outline possible strands of further
research in RISCAL and SMT technology. \autoref{sect:appendix} includes
detailed illustrations of the benchmark results. More details can be found
in the technical report~\cite{Schreiner2021} on which this paper is based.
Due to space restrictions, we have to refer the reader to~\cite{RISCAL} for
an overview on the RISCAL language which is elucidated by a tutorial and
reference manual and various publications.

\clearpage

\section{Deciding First-Order Formulas}
\label{sect:decide}

In the following, we briefly describe the two alternative mechanisms that
RISCAL implements for deciding first-order formulas: internal semantic
evaluation and the application of external SMT solvers.

\paragraph{Semantic Evaluation}

The built-in decision mechanism of RISCAL is based on the translation of
every syntactic phrase of the RISCAL language into an executable
representation of its denotational semantics. This representation is a Java
\enquote{lambda expression} that in essence maps an assignment for the free
variables of the phrase to the value denoted by its semantics, i.e., the
truth value of a formula or the updated variable assignment resulting from
the execution of a command~\cite{Schreiner2020c}. In the case of first-order
logic formulas, the most interesting part of the translation is that of a
universally quantified formula $\symforall{\symtype{x}{D}}{F}$ and that
of an existentially quantified formula $\symexists{\symtype{x}{D}}{F}$,
respectively; these translations are semi-formally sketched below (here
$\sem{F}$ denotes the body of a function whose execution yields the truth
value of $F$):

\begin{minipage}{\textwidth}\small
\begin{minipage}{0.49\textwidth}
\begin{align*}
& \sem{\symforall{\symtype{x}{D}}{F}} \symdeff \\ 
& \quad e\ \progtext{:= enumerate($D$)} \\[-0.25em]
& \quad\progtext{loop} \\[-0.2em]
& \qquad \progtext{if empty($e$) then return true} \\[-0.25em]
& \qquad \progtext{$x$ := next($e$); $e$ := rest($e$)} \\[-0.25em]
& \qquad \progtext{if $\neg\texttt{call}(\sem{F},x)$ then return false}\\[-0.25em]
\end{align*}
\end{minipage}
\begin{minipage}{0.49\textwidth}
\begin{align*}
& \sem{\symexists{\symtype{x}{D}}{F}} \symdeff \\ 
& \quad e\ \progtext{:= enumerate($D$)} \\[-0.25em]
& \quad\progtext{loop} \\[-0.2em]
& \qquad \progtext{if empty($e$) then return false} \\[-0.25em]
& \qquad \progtext{$x$ := next($e$); $e$ := rest($e$)} \\[-0.25em]
& \qquad \progtext{if \texttt{call}(\sem{F},x) then return true}\\[-0.25em]
\end{align*}
\end{minipage}
\end{minipage}

The core of the translation is a loop that enumerates every element
of the domain~$D$ of the quantified variable~$x$ and evaluates the body of
the quantified formula with $x$ bound to that element, until the truth value
of the body determines the overall result. As an optimization, RISCAL
actually implements the enumeration of $D$ in a mostly \enquote{lazy}
fashion such that it is not necessary to simultaneously keep all elements in
memory; the generation stops when the first element has been produced that
allows to decide the formula. Consequently, the \enquote{worst case} is
exhibited by a \emph{true universal formula} or a \emph{false
existential formula}: here we have to generate all elements, before we can
decide that the universal formula is true or the existential
formula is false.

Furthermore, RISCAL supports expressions that do not denote unique values,
for example the term $(\textsf{choose}\ \symtype{x}{D}\ \textsf{with}\
F[x])$ that denotes any value $x$ of the domain~$D$ that satisfies the
formula~$F[x]$. RISCAL implements such a term in its
\enquote{nondeterministic} evaluation mode~\cite{Schreiner2020c} by the
computation of a (lazily evaluated) \emph{stream} of such values. Like for
quantified formulas, the core of this translation is a loop that enumerates
every element of $D$; the translation yields each element that satisfies the
body formula as a value of the expression (i.e., this value is appended to a
stream of values denoted by the term). A formula that depends on such terms
correspondingly denotes a stream of truth values; the formula is only
considered as valid if this stream only consists of instances of truth
value~\enquote{true}.
Not necessarily unique choices arise in many mathematical definitions and
algorithms (\enquote{choose any element $e$ of set $S$}). Furthermore,
applications of such expressions may emerge from the modular verification of
user-defined operations; here not the definition of an operation but its
contract is considered. For instance, an application~$f(a)$ of a function
$f$ specified as
\begin{align*}
&\symc{fun}\ f(\symtype{x}{D})\colon{D}\ \textsf{ensures}\ F[x,\mathit{result}]
\end{align*}
(where $\mathit{result}$, is a special variable that denotes the result of
the function) can be replaced by the expression
$(\symchoose{\symtype{\symv{result}}{D}} F[a,\symv{result}])$.

\paragraph{SMT Solving}

The problem of deciding the validity of a formula $F$, denoted as 
$\mathit{valid} \sem{F}$, can be reduced to the problem of deciding
the satisfiability of the negation of $F$, denoted as
$\mathit{sat} \sem{\neg F}$, by applying the equivalence.
\(
\mathit{valid} \sem{F} \equiv \neg \mathit{sat} \sem{\neg F}.
\)
RISCAL implements a translation to formulas in the
SMT-LIB format~\cite{SMTLIB}; thus we can decide the validity of the RISCAL
formula $F$ by letting an external SMT solver decide the satisfiability of
the SMT-LIB version of $\neg F$. As a background theory we have chosen the
theory of \emph{fixed-size bit vectors}: since every RISCAL domain is
finite, every element of a domain with $n$ elements can be represented by a
vector of $\lceil \log n\rceil$ bits; furthermore the set of bit vector
operations is expressive enough to allow a proper
encoding of the various RISCAL operations. However, bit vectors alone are
not enough: the treatment of quantifiers and choose expressions (discussed
below) requires functions which are not explicitly characterized by
definitions but only implicitly by axioms; therefore we demand from the
theory also support for \emph{uninterpreted functions}. Furthermore, the main
SMT-LIB logic that provides bit vectors and uninterpreted function is the
logic~\textrm{QF\_UFBV} of \enquote{unquantified formulas over bit vectors
with uninterpreted sort and function symbols} which is supported, e.g., by
the well known SMT solvers Boolector, CVC4, Yices, and Z3. We therefore have
to translate a RISCAL formula with quantifiers into a corresponding
quantifier-free SMT-LIB formula (since some SMT solvers actually support as a
non-standard extension also quantified bit vector formulas with
uninterpreted functions, we will in the benchmarks later also experiment
with the preservation of quantifiers).

The problem of eliminating quantifiers is addressed by the following
equivalences which semi-formally sketch how quantifiers can be removed from
RISCAL formulas (the role of function~$f$ is explained below):
\begin{align*}
\mathit{valid}\sem{\symexists{\symtype{x}{D}}{F[x]}} 
&\equiv \symnot \mathit{sat}\sem{\symnot\symexists{\symtype{x}{D}}{F[x]}} \\
&\equiv \symnot \mathit{sat}\sem{\symforall{\symtype{x}{D}}{\symnot F[x]}} \equiv
\symnot \mathit{sat}\sem{\symnot F[e_1]\symand\ldots\symand \symnot F[e_n]} \\[0.5em]
\mathit{valid}\sem{\symforall{\symtype{x}{D}}{F[x]}}
& \equiv \symnot \mathit{sat}\sem{\symnot\symforall{\symtype{x}{D}}{F[x]}} \\
& \equiv \symnot \mathit{sat}\sem{\symexists{\symtype{x}{D}}{\symnot F[x]}} \equiv
\symnot \mathit{sat}\sem{\symnot F[f(x_1,\ldots,x_n)]} 
\end{align*}
We assume that before the translation is applied all formulas have been
transformed into \emph{negation normal form}, i.e., all applications of
the negation symbol have been \emph{pushed inside} down to the level 
of atomic formulas. Thus, above occurrences of quantified formulas are
\emph{positive}, i.e., they do not appear in the context of negation.
Then the decision of $\mathit{valid}\sem{\symexists{\symtype{x}{D}}{F[x]}}$
boils down to the decision of the satisfiability of the universally
quantified formula $\symforall{\symtype{x}{D}}{\symnot F[x]}$. Now, if $D$
consists of $n$ values denoted by terms $e_1,\ldots,e_n$, we can expand the
quantified formula to an equivalent conjunction $\symnot
F[e_1]\symand\ldots\symand \symnot F[e_n]$. On the other hand, the decision of
$\mathit{valid}\sem{\symforall{\symtype{x}{D}}{F[x]}}$ boils down to the
decision of the satisfiability of the existentially quantified formula
$\symexists{\symtype{x}{D}}{\symnot F[x]}$. Analogously to the previous
case, we could in principle also expand this formula, namely to a 
disjunction $\symnot F[e_1]\symor\ldots\symor\symnot F[e_n]$.
However, for this kind of decision we generally prefer another option that avoids the 
blow-up of the formula. Let us assume that the existentially quantified formula appears in
the context of $n$ universally quantified variables $x_1,\ldots,x_m$, i.e.,
the problem of deciding the satisfiability of
$\symexists{\symtype{x}{D}}{\symnot F[x]}$ actually occurs in the course of
deciding the satisfiability of a global formula of the shape
\(
\symforall{\symtype{x_1}{D_1}}\ldots\symforall{\symtype{x_m}{D_m}}\ldots
\symexists{\symtype{x}{D}}{\symnot F[x]}
\).
Then we introduce an $m$-ary function symbol $f$ that does
not appear anywhere else in the global formula; the denoted function
can therefore have an arbitrary interpretation (we call such a function a
\emph{Skolem function}). Finally, we replace $\symexists{\symtype{x}{D}}{\symnot
F[x]}$ by $\symnot F[f(x_1,\ldots,x_m)]$. In the special case $m=0$, i.e., if
there is no outer universally quantified variable, $f$ becomes a
\emph{Skolem constant} and the formula becomes $\symnot F[f]$.
Although the resulting formula is not logically equivalent to the original
one, it is \emph{equi-satisfiable}, i.e., it is satisfiable if and only
if the original formula is (if we may choose for all values $x_1,\ldots,x_m$
a value for~$x$ that makes $F[x]$ true, then from these choices we may
construct the Skolem function~$f$ and vice versa). Since the translation
preserves satisfiability, the equivalence stated above holds.

From the above translation, deciding the validity of an existentially
quantified formula may blow-up the formula to a size that is exponential in
the depth of the nesting of existential quantifiers; this may also increase
the complexity of the decision. Furthermore, problems may arise even with
the decision of the validity of a universally quantified formula, which
entails deciding the satisfiability of a formula $\symnot
F[f(x_1,\ldots,x_m)]$ with Skolem function~$f$ (please note that
$x_1,\ldots,x_m$ represent concrete values, the translated formula does not
have any free variables). In the translation to the SMT-LIB theory
\textrm{QF\_UFBV}, the range of $f$ is not anymore the original RISCAL
domain~$D$ of the existentially quantified variable, but some bit vector
type $B$ whose values encode the values of~$D$. Since not every bit vector
in~$B$ necessarily represents an element from~$D$, we have to constrain the
range of $f$ by a predicate~$p_D$ that holds for a bit vector $b \in B$
if and only if $b$ actually represents an element from~$D$. We achieve this
by adding to the translation an axiom
\[
\bigwedge_{d_1,\ldots,d_m} p_D(f(t(d_1),\ldots,t(d_m)))
\]
where $t(d)$ represents an expression that denotes the bit vector associated to the RISCAL
value $d$. The size of this conjunction is proportional to the number of
possible combinations of values $d_1,\ldots,d_m$ for the arguments of $f$;
this may blow up the SMT-LIB translation considerably and overcome the
benefits of applying Skolemization rather than expansion. Thus the
translation can be configured to apply a heuristic: if the number
of conjuncts in the Skolemization axiom is significantly larger than the
number of conjuncts derived from expanding the original formula, the
translation forsakes Skolemization in favor of expansion.

While expressions denoting unique values can be directly encoded by bit
vectors operations, an
$(\symchoose{\symtype{y}{D}}{F[x,y]})$ with free variable $\symtype{x}{D}$
gives in the SMT-LIB translation rise to a new function $f\colon D\to D$
with axiom $\symforall{\symtype{x}{D}} {F[x,f(x)]}$. Similarly, in modular verification, every RISCAL
operation (function, predicate, procedure) specified by a contract gives
rise to an SMT-LIB function with a corresponding axiomatization. As
explained above, such axiomatizations by universally quantified formulas
yield large SMT-LIB expansions and potentially costly SMT decisions (in
addition to the user-defined axiomatization, such functions have also to be
constrained by the type representation axioms explained above).
However, in certain contexts we may replace applications of
such axiomatized functions. For instance, take the formula
\(
\forall a.\ (\ldots f(a) \ldots)
\)
where application $f(a)$ occurs positively (unnegated) in a context $(\ldots\
f(a) \ldots)$ that does not embed $f(a)$ in another quantifier and
assume that function $f\colon D\to D$ has been axiomatized as 
described above. Then above formula can be transformed to the 
equi-satisfiable formula \(
\forall a,b.\ ({F[a,b]} \symimp \ldots b \ldots).
\)
Thus, we have replaced the application $f(a)$ of axiomatized function~$f$ by a
fresh universally quantified variable $b$ with assumption $F[a,b]$. This
means that the original axiomatization (which applied formula~$F$ to
arbitrary values $x$ from $D$) has been specialized to the instances that
are actually relevant. RISCAL optionally implements in the SMT-LIB translation a
generalized form of this transformation under the name \enquote{eliminate
choices}, because it is directly applied to choose expressions given by the
user and to choose expressions generated from applications of implicitly
defined functions. In combination with the option \enquote{inline
definitions}, also choose expressions indirectly arising from the
definitions of operations may be inlined. While this expands the size of the
core formula to be decided, it removes general axiomatizations and may
thus be beneficial all in all.

\paragraph{Comparison}

As discussed above, deciding by SMT solving formulas with quantifiers or
choose expressions can become problematic, because the SMT-LIB translation
may yield vastly expanded formulas (arising from existential formulas,
quantified constraints of Skolem functions, and axioms of uninterpreted
functions emerging from choose expressions or modular
verification). To which extent this affects the actual performance of the
decision process can be investigated only by actual benchmarks.

\section{Artificial Benchmarks}
\label{sect:artificial}

\paragraph{Basic Setup}

We start by investigating the \enquote{base behavior} of the two decision
approaches. For this, we use the following two predicates:
\begin{align*}
\textit{cycle4-valid} 
&\equiv \neg (x_1 < x_2 \wedge x_2 < x_3 \wedge x_3 < x_4 \wedge x_4 < x_1) \\
\textit{cycle4-sat1} 
&\equiv \neg (x_1 < x_2 \wedge x_2 < x_3 \wedge x_3 < x_4 \wedge x_4 < x_1+4)
\end{align*}
Both predicates have free occurrences of four integer variables~$x_1,x_2,x_3,x_4$.
Predicate~\textit{cycle4-valid} states that these variables
cannot form a \enquote{less-than cycle}; this predicate is valid
and its negation is unsatisfiable. On the other side, predicate~\textit{cycle4-sat1} is
satisfiable but not valid, as is its negation. However, while
\textit{cycle4-sat1} has many satisfying assignments, its negation
has only few; thus~\textit{cycle4-sat1} represents a \enquote{mostly
valid} formula, while its negation denotes a \enquote{mostly unsatisfiable}
one. Both predicates only depend on the atomic predicate~$<$ (the second one
also on the constant addition~$+4$). The predicates
do not require any complex calculations or decisions in order to
most clearly exhibit the effect of various forms of quantification
structures on the decision process.
Here we investigate the eight quantification patterns
$\exists^4\forall^0$, $\exists^3\forall^1$,$\exists^2\forall^2$,
$\exists^1\forall^3$, $\forall^4\exists^0$, $\forall^3\exists^1$,
$\forall^2\exists^2$, $\forall^1\exists^3$ where $Q^i$ represents the
$i$-fold repetition of quantifier~$Q$ and the variables are quantified in
the order $x_1,x_2,x_3,x_4$. Thus, e.g., the combination of quantification
pattern $\exists^3\forall^1$ with
predicate \textit{cycle4-valid} represents the following formula:
\begin{align*}
\exists \symtype{x_1}{D}, \symtype{x_2}{D}, \symtype{x_3}{D}.\
\forall \symtype{x_4}{D}.\ \neg (x_1 < x_2 \wedge x_2 < x_3 \wedge x_3 < x_4 
\wedge x_4 < x_1)
\end{align*}
Above quantifier patterns consider the cases of purely existential formulas
($\exists^4\forall^0$), purely universal formulas ($\forall^4\exists^0$), as
well as existential formulas with universal bodies ($\exists^i\forall^j$) and
universal formulas with existential bodies ($\forall^j\exists^i$) where the
number of corresponding quantifiers represent different sizes of the
respective quantification ranges. 
As for the domain~$D$ of the variables, we focus on $D:=\mathbb{N}[2^N-1]$
for some $N\in\mathbb{N}$, i.e., each of the 4 variables holds some natural
number up to maximum~$2^N-1$; the total value space thus consists of
$2^{4N}$ elements. In the following benchmarks, we choose $N:=6$, i.e., a
value space of size $2^{24}$. For a formula of shape $\exists^i\forall^j$
or $\forall^j\exists^i$, this value space is partitioned according
to the numbers~$i$ and $j$ of existentially and universally quantified
variables, respectively. The \enquote{existential search space} has size
$2^{iN}$, which leads in the QF\_UFBV translation to the generation of
$2^{iN}$ clauses. The \enquote{universal search space} has size $2^{jN}$,
which leads in QF\_UFBV to $j$ Skolem constants (if the universal
quantifiers are outermost) or $j$~Skolem functions of arity $i$ (if the
universal quantifiers are innermost); the domain of each Skolem constant or
function is a bit vector of length~$N$ with $2^N$ possible values.

\paragraph{Experimental Results}

\begin{figure}
\begin{center}
\includegraphics[width=0.48\textwidth]{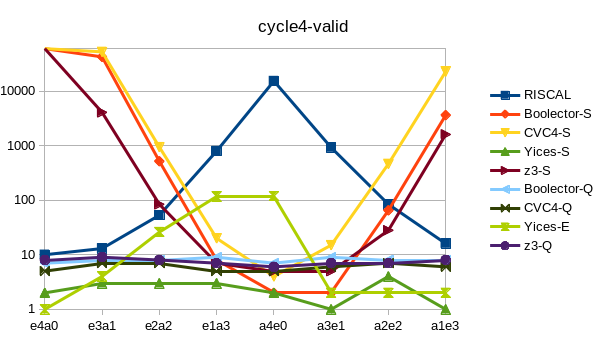} \quad
\includegraphics[width=0.48\textwidth]{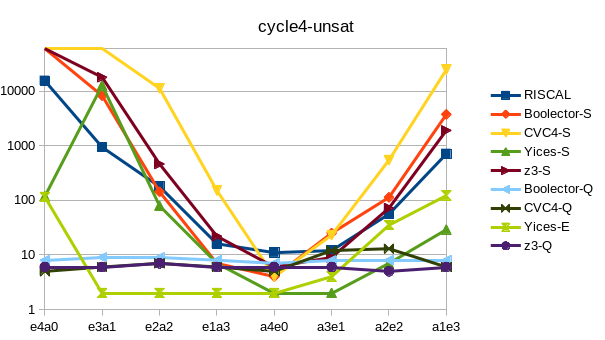} \\
\includegraphics[width=0.48\textwidth]{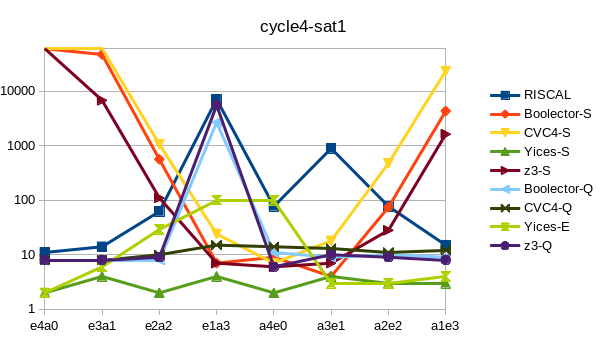} \quad
\includegraphics[width=0.48\textwidth]{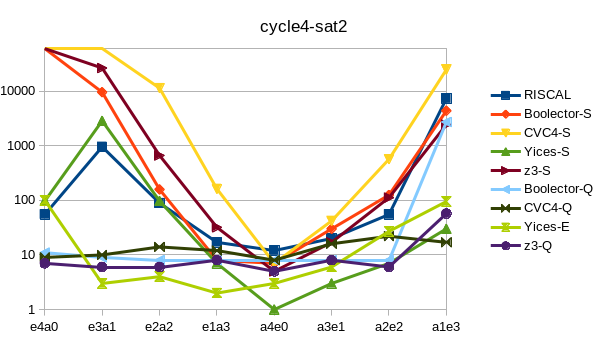} 
\end{center}
\caption{Artificial Benchmarks: Base Behavior}
\label{fig:base}
\end{figure}

The four diagrams in \autoref{fig:base} plot the decision times for the
quantified formulas with (valid) predicate \textit{cycle4-valid} and its
(unsatisfiable) negation and for the satisfiable (mostly valid) predicate
\textit{cycle4-sat1} and its also satisfiable (but mostly unsatisfiable)
negation. The labels of the horizontal axis denote the applied
quantification pattern (labels~\textsf{e$i$a$j$} and \textsf{a$i$e$j$}
denote patterns $\exists^i\forall^j$ and $\forall^i\exists^j$,
respectively). The vertical axis denotes the decision time in ms, within the
interval $[1,60000]$ (please note the logarithmic scale). All decision
procedures were forcefully terminated after 1~minute; thus, if a plot point
is at the top line of the diagram, this actually indicates \enquote{timeout}
or \enquote{no result} (a timeout is also indicated, if the software ran out
of memory or produced any other kind of error). All measurements were
performed on a virtual GNU/Linux machine with a CPU of type
i7-2670QM@2.20GHz using 8~GB RAM. In case of the SMT solvers, only the time
for the actual decision (not including the time for translating the RISCAL
formula to an SMT-LIB formula) was considered.

The various labeled curves give the times for the decision mechanisms that
we have benchmarked using RISCAL 3.8.5 and the SMT solvers Boolector~3.1.0,
CVC4~1.7, Yices~2.6.1, and Z3~4.8.7. Here tag \textsf{RISCAL} represents the
built-in semantic evaluation mechanism of RISCAL. Tags \textsf{Boolector-S},
\textsf{CVC4-S}, \textsf{Yices-S}, \textsf{Z3-S} represent the application
of the various SMT solvers to the generated QF\_UFBV formula where
quantifiers are removed by Skolemization (in case of the originally
universal quantifiers) or expansion (in case of the existential
quantifiers). Tags \textsf{Boolector-Q}, \textsf{CVC4-Q}, \textsf{Z4-Q}
represent the application of the SMT solvers where, however, in the
generated formula all quantifiers are preserved (Boolector, CVC4, and Z3
also support quantification). \textsf{Yices-E} represents the application of
Yices where also the (originally) universal quantifiers have been removed by
expansion rather than by Skolemization. Thus every SMT solver is benchmarked
twice, with two different mechanisms for dealing with quantifiers: by
eliminating them as described in \autoref{sect:decide} to yield a formula in
the standard logic QF\_UFBV of SMT-LIB, or by applying the non-standard
quantification support of the various SMT solvers. Only in the case of Yices
(which has only a limited support for quantification), we apply the
alternative of expanding also (originally) universal quantifiers.

An inspection of the diagrams shows that, when eliminating quantifiers by
Skolemization or expansion, Yices is mostly the fastest among the
benchmarked SMT solvers; the other solvers are able to compete with Yices
only if quantifiers are preserved in the formulas. The semantic evaluation mechanism is mostly outperformed by Yices
and also by the other solvers. However, the other solvers are superior only
if the quantifiers are preserved in the formulas, or if we consider the
cases in the middle of the left diagrams (few or no existential quantifiers
and mainly valid base predicates). When quantifiers are eliminated, the
semantic evaluation mechanism of RISCAL outperforms Boolector, CVC4, and
Yices at the boundaries of the left diagrams; also in the right diagrams (mostly unsatisfiable base predicates), the
performance of semantic evaluation at least matches that of the solvers.

Furthermore, the semantic evaluation mechanism of RISCAL exhibits
comparatively good performance in diagram~\textit{cycle4-valid} for the
quantification pattern~$\exists^i\forall^j$. Since the outermost quantifier
is existential, only a single value for its variable has to be found that
makes the formula true; since the base predicate is valid, already the first
choice is successful. The more existential quantifiers follow, i.e., the
bigger~$i$ is, the bigger the advantage is. If all quantifiers are
existential (case~$\exists^4\forall^0$), the first attempted choice for all
variables already leads to a decision of the formula. However, if more and
more variables get universally quantified, the more and more work has to be
performed to validate the existential choice. The worst situation arises, if
all variables are universally quantified (case~$\forall^4\exists^0$); here
the full variable space has to be investigated to determine the validity of
the formula. However, the more of the inner variables get existentially
quantified, the quicker the decision for each value of a universally
quantified variable becomes. If only the outermost variable is universally
quantified (case~$\forall^1\exists^3$), only the space of the outermost
variable has to be fully investigated. This explains the shape of the RISCAL
curve which grows from the fully existentially quantified formula of type
$\exists^4\forall^0$ until it reaches a sharp peak at the fully universally
quantified formula of type $\forall^4\exists^0$; then the curve goes down
again towards the formula pattern~$\forall^1\exists^3$. On the other hand,
plot~\textit{cycle4-unsat} illustrates the dual behavior for the negated
(unsatisfiable) version of the predicate. To show that the fully
existentially quantified formula $\exists^4\forall^0$ is false, the whole
value space has to be investigated, while for the fully universally
quantified formula $\forall^4\exists^0$ the first encountered value
combination represents a counterexample to the truth of the formula; for a
growing number of inner existential quantifiers, again more value
combinations have to be investigated, though. Finally, the two plots for the
mostly satisfiable predicate \textit{cycle4-sat1} and its mostly
unsatisfiable negation \textit{cycle4-sat2} are similar to the plots for the
valid and unsatisfiable cases, except that there is no more a pronounced
\enquote{peak} (maximum or minimum) for the fully universally quantified
pattern $\forall^4\exists^0$: the investigation of the value space can stop
when the first counterexample is found, but not necessarily the first value
combination encountered immediately represents such a counterexample.

More results are given in \autoref{sect:appendix} which illustrates in
\autoref{fig:valid} and \autoref{fig:sat} corresponding benchmarks with more
complex predicates involving operations such as non-linear arithmetic,
arithmetic quantification, set and array operations. These benchmarks reveal
various situations when the semantic evaluation mechanism of RISCAL is able
to compete with or even outperform some of the SMT solvers; for a more
detailed interpretation of these and other benchmarks,
see~\cite{Schreiner2021}.

So far, we have only considered formulas with built-in operations (functions
and predicates). Now we are going to also consider operations
specified by contracts such as the following two functions:
\begin{align*}
& \textsf{fun}\ f(x_1,x_2,x_3,x_4)\ \textsf{ensures}\\
& \quad\textsf{if}\ x_1 < x_2 \wedge x_2 < x_3 \wedge x_3 < x_4 \wedge x_4 < x_1\
\textsf{then}\ \mathit{result} = 0\ \textsf{else}\ \mathit{result} = 1; \\
& \textsf{fun}\ g(x_1,x_2,x_3,x_4)\ \textsf{ensures}\\
& \quad\textsf{if}\ x_1 = x_2 \wedge x_3 = x_4 \
\textsf{then}\ \mathit{result} = 0\ \textsf{else}\ \mathit{result} = 1; 
\end{align*}
Here $f(x_1,x_2,x_3,x_4)$ is 1 for all $x_1,x_2,x_3,x_4$ and
$g(x_1,x_2,x_3,x_4)$ may be 1 or 0, depending on the values
of $x_1,x_2,x_3,x_4$. We will consider
quantified formulas with the quantification patterns used in
the previous sections, using four base predicates that test the equality
of above functions with values~$1$ or $0$, yielding again
one valid, one unsatisfiable, and two satisfiable situations.

\begin{figure}
\begin{center}
\includegraphics[width=0.48\textwidth]{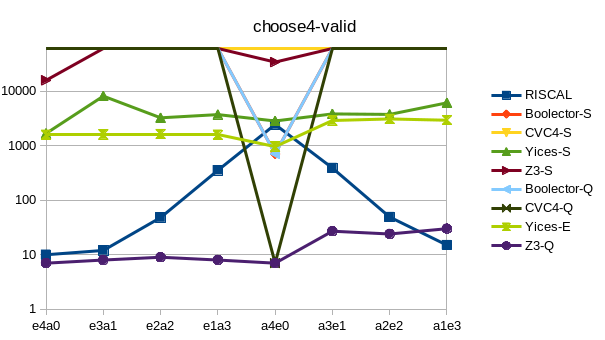} \quad
\includegraphics[width=0.48\textwidth]{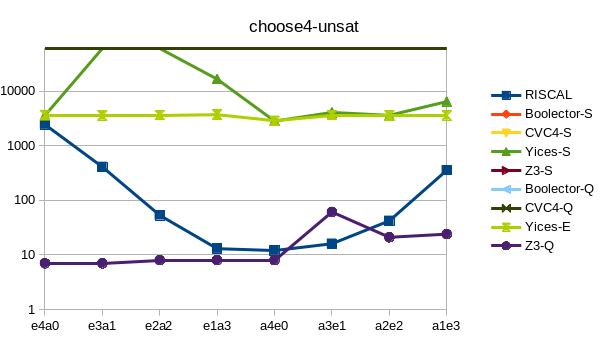} \\
\includegraphics[width=0.48\textwidth]{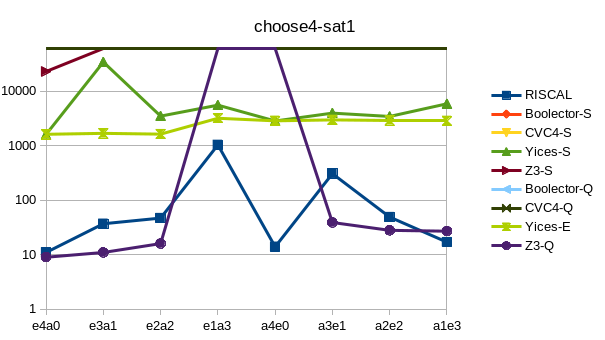} \quad
\includegraphics[width=0.48\textwidth]{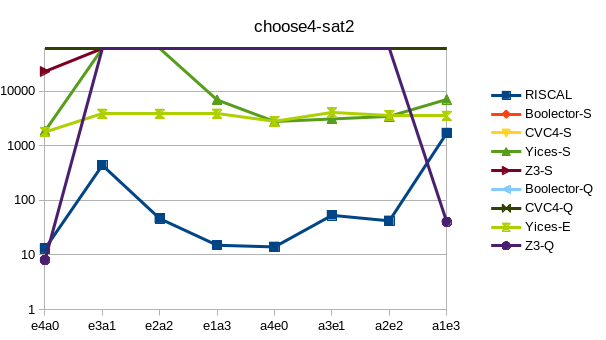} 
\end{center}
\caption{Formulas with Functions Specified by Contracts}
\label{fig:choose}
\end{figure}

\autoref{fig:choose} displays the decision times for model parameter $N:=5$
(we now use value~$5$ rather than $6$ to compensate the additional quantifier
of the function axiom). Here the evaluation mechanism of RISCAL is generally
faster than the SMT solvers (indeed Boolector and CVC4 do mostly not deliver
any answers within the given time bound); the major exception is Z3 if
quantifiers are preserved, which is faster than RISCAL for valid and
unsatisfiable formulas. Yices also produces results but is mostly much
slower than RISCAL. Boolector does not support uninterpreted functions if
quantifiers are preserved, thus also the benchmark set \enquote{Boolector-Q}
expands (as \enquote{Boolector-S} does) universal and existential
quantifiers to conjunctions and disjunctions, respectively.
	
The results demonstrate that uninterpreted functions characterized by axioms
rather than definitions pose a major problem for most SMT solvers; in the
presence of such functions often the nondeterministic evaluation mechanism
of RISCAL is superior. In an attempt to mitigate this problem a bit, we have
applied the previously mentioned options \enquote{eliminate choices} and
\enquote{inline definitions} to eliminate certain applications of
contract-specified operations by embedding the postconditions into the
enclosing formulas; this does not change the satisfiability of the formula, if
the application occurs in a non-negated purely universally quantified
context. In our experiments, this is the case (only) for the quantifier
pattern \textsf{a4e0} ($\forall^4\exists^0$) which indeed shows substantial
speedups (only) with the application of Yices; the experimental results
given above have been derived with these options.

\section{Real-Life Benchmarks}
\label{sect:real}

The artificial benchmarks investigated so far do not demonstrate whether/how
often the observed effects indeed emerge in \enquote{real-life} examples. To
shed some light on this issue, we have also collected from real RISCAL
models a selection of formulas whose validity is to be decided. These
formulas mainly represent conditions to validate the specifications of
problems, verify the correctness of algorithms, or also theorems over
the domains of consideration. In the overwhelming majority of cases, the
decision of such conditions by SMT solving vastly outperforms the decision
by semantic evaluation. However, for the purpose of this paper, we have
explicitly selected a sample where this is not the case, i.e., where
semantic evaluation is competitive with SMT solving.

\begin{figure}
\centering
\includegraphics[width=0.99\textwidth]{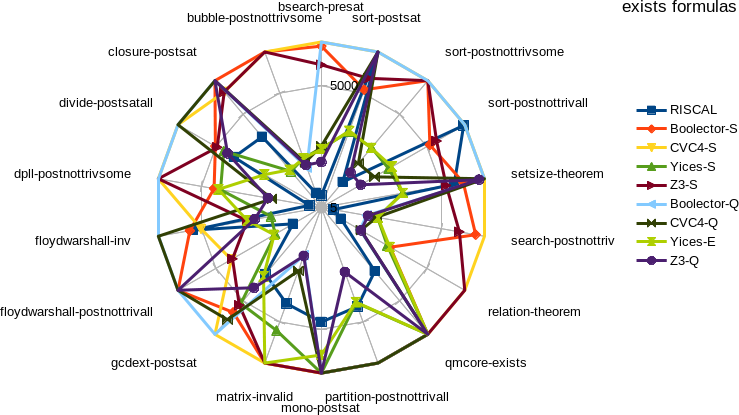} 
\caption{Real-Life Benchmarks}
\label{fig:real1}
\end{figure}

The diagram displayed in \autoref{fig:real1} presents the results of
benchmarking the decision of some of these formulas
(see~\autoref{sect:appendix} or~\cite{Schreiner2021} for a link to
the sources). The benchmarks are visualized as a circular \enquote{net}
chart where each radial represents one formula whose validity is to be
decided, the center represents some minimal time and the outermost rim of
the net represents the 60 seconds timeout limit. Therefore, the closer the
lines connecting the benchmark points of a particular decision mechanism are
to the center of the net, the faster the decision mechanism is; a line along
the outermost rim indicates timeout situations. As in the artificial
benchmarks, the values along the radials are plotted in logarithmic
magnitudes. In particular, the diagram illustrates benchmarks for formulas that have substantial
\enquote{existential} content; many stem from
validating procedure contracts by checking the satisfiability of the
postconditions for all inputs that satisfy the preconditions; here 
the semantic evaluation of RISCAL often outperforms the
various SMT solvers. The best SMT results are achieved by Z3 when preserving
quantifiers (\textsf{Z3-Q}) and Yices with either Skolemization or expansion
of the (original) universal quantifiers (\textsf{Yices-S} and
\textsf{Yices-E}). 

\autoref{sect:appendix} lists in \autoref{fig:real2} more benchmarks where
the left column illustrates the execution of the benchmarks with smaller
values for the model parameters, while
the right column illustrates executions with larger
values; all executions were again terminated after 60 seconds. The top row
gives the benchmarks already shown in \autoref{fig:real1}, but now also with
larger model parameters. The second row illustrates benchmarks for formulas
with choose expressions; here in the SMT decisions the option
\enquote{choose elimination} transformation was \emph{not} applied. The
evaluation mechanism of RISCAL again outperforms many of the SMT solvers;
among these typically Yices performs best. The third row illustrates
benchmarks for the same formulas as in the second row but with the options
\enquote{choose elimination} and \enquote{inline definitions} turned on.
This shows a clear improvement in many examples; now various SMT
solvers clearly beat the semantic evaluation mechanism of RISCAL. The last
row illustrates benchmarks for formulas that do not obviously fall
into above categories but where nevertheless the semantic evaluation
mechanism is competitive, from the structure of the formulas
and/or the complexity of the underlying operations.

\clearpage

\section{Conclusions}
\label{sect:concl}

Generally the decision of first-order formulas over finite domains by
external SMT solvers via a translation into the SMT-LIB
logic~\textrm{QF\_UFBV} (unquantified formulas over bit vectors with
uninterpreted sort and function symbols) and applying external SMT solvers
vastly outperforms the semantic evaluation mechanism built-in into RISCAL.
In this paper, however, we have also identified cases where this is not
necessarily the case, mainly because the SMT-LIB translation leads to a large
number of clauses in the generated conjunctive normal form.

One case are theorems with substantial \enquote{existential} content, i.e.,
positive occurrences of existential quantifiers with large quantification
ranges: the resulting formulas may be so big that their decision by SMT
solving may be outperformed by semantic evaluation. Another case are
theorems with uninterpreted functions that are axiomatized by universally
quantified formulas with large quantification ranges; also these axioms lead
to the generation of SMT-LIB formulas with a huge number of clauses that
slow down the execution of SMT solvers. This problem may be partially
mitigated, if applications of such functions occur in a pure universal
context; an optimization technique may replace the function application by
universally quantified variables that are constrained by an appropriate
instance of this axiom.

Furthermore, also for universally quantified theorems the problem arises
that the Skolem functions generated from their negated counterparts have to
be constrained by axioms that describe which bit vector values indeed
denote valid RISCAL values. RISCAL therefore implements an SMT-LIB option
that applies a heuristic to decide whether it is cheaper to expand the
quantified formula rather than to generate a Skolem function. Another
problem may result from the necessary encoding of various operations on the
data types supported by RISCAL (such as non-linear arithmetic, arithmetic
quantifiers, or set size computations) where the built-in evaluation
mechanisms of RISCAL may perform better than the corresponding RISCAL
encodings; however, while this effect can be observed in artificial
benchmarks, it seems not to be a major problem in most real-life examples.
	

Our work demonstrates that there is still room for improvement in current
SMT solvers for the SMT-LIB logic~\textrm{QF\_UFBV} with respect to deciding
theorems with substantial existential content and applications of
axiomatized functions. On the other side, we will continue to investigate
how the translation of RISCAL formulas to SMT-LIB formulas can be optimized
to take the presented findings into account.

%% file: appendix.tex
\appendix
\section{Benchmark Diagrams}
\label{sect:appendix}

\begin{figure}[H]
\begin{center}
\includegraphics[height=0.1465\textheight]{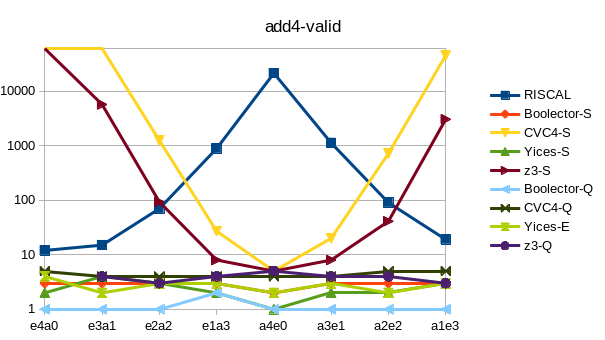} \quad
\includegraphics[height=0.1465\textheight]{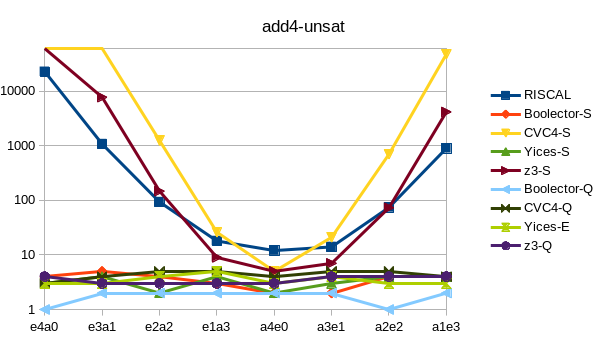} \\
\includegraphics[height=0.1465\textheight]{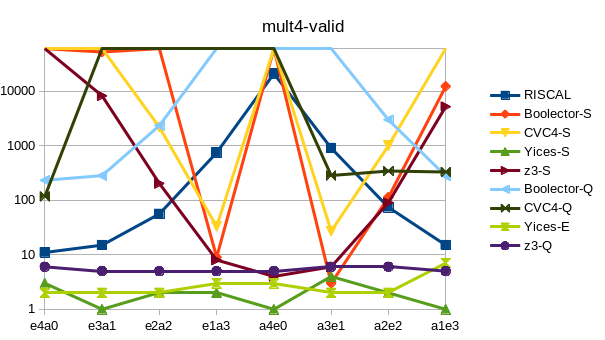} \quad
\includegraphics[height=0.1465\textheight]{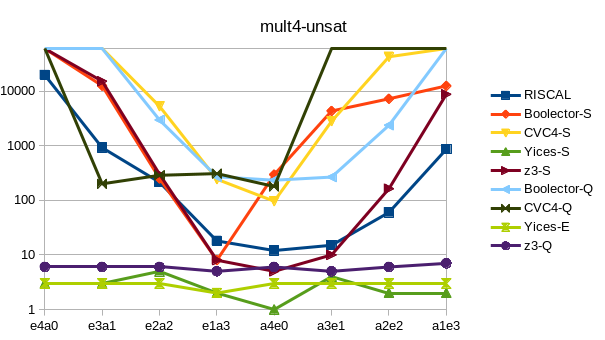} \\
\includegraphics[height=0.1465\textheight]{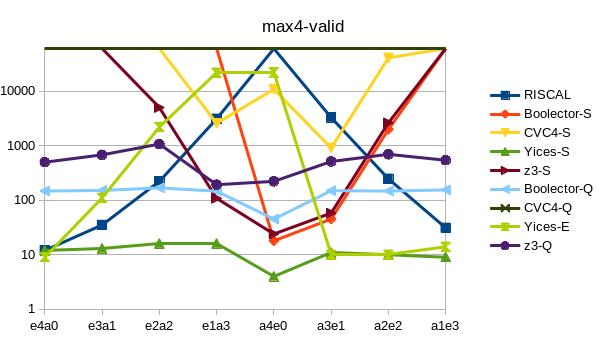} \quad
\includegraphics[height=0.1465\textheight]{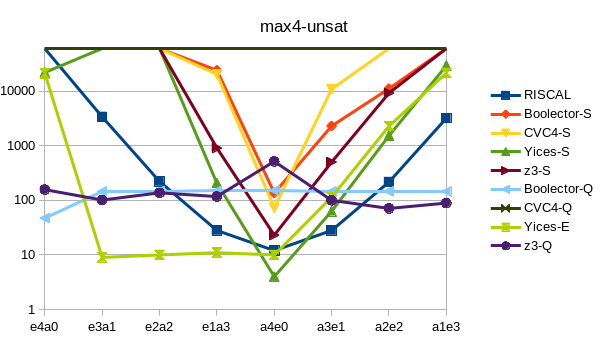} \\
\includegraphics[height=0.1465\textheight]{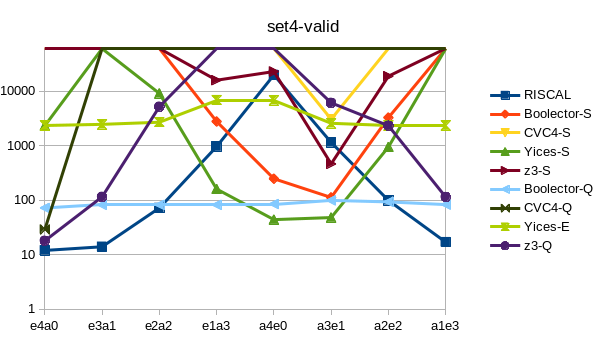} \quad
\includegraphics[height=0.1465\textheight]{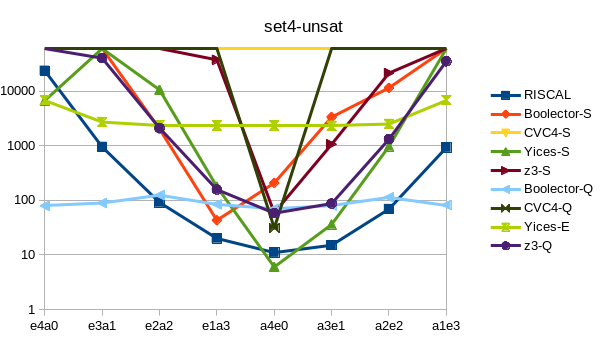} \\
\includegraphics[height=0.1465\textheight]{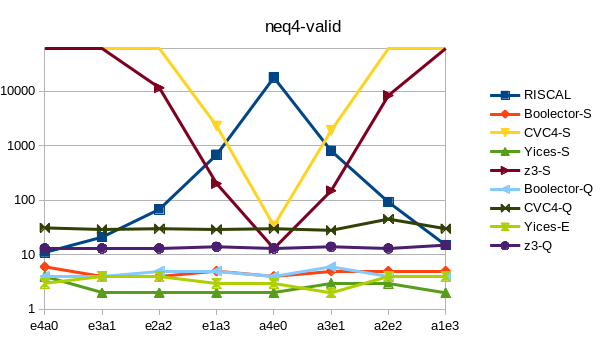} \quad
\includegraphics[height=0.1465\textheight]{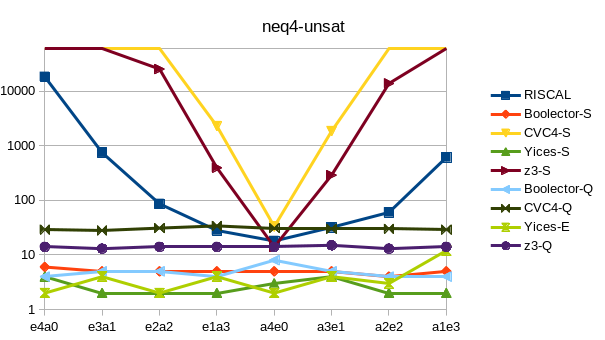} \\
\includegraphics[height=0.1465\textheight]{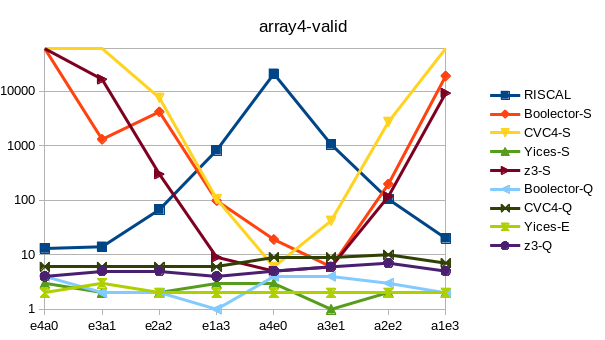} \quad
\includegraphics[height=0.1465\textheight]{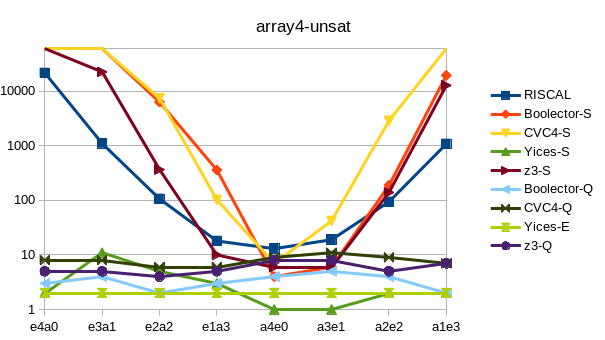} 
\end{center}
\caption{Artificial Benchmarks: Valid versus Unsatisfiable Predicates}
\label{fig:valid}
\end{figure}

\begin{figure}[H]
\begin{center}
\includegraphics[height=0.155\textheight]{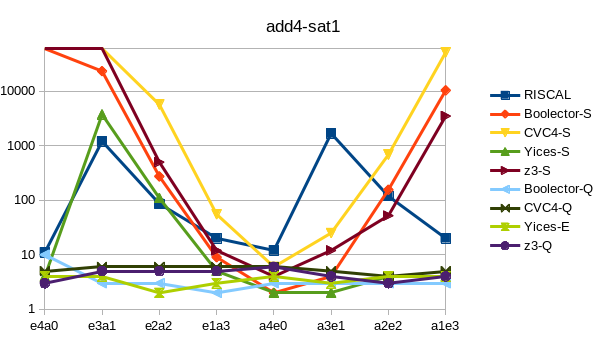} \quad
\includegraphics[height=0.155\textheight]{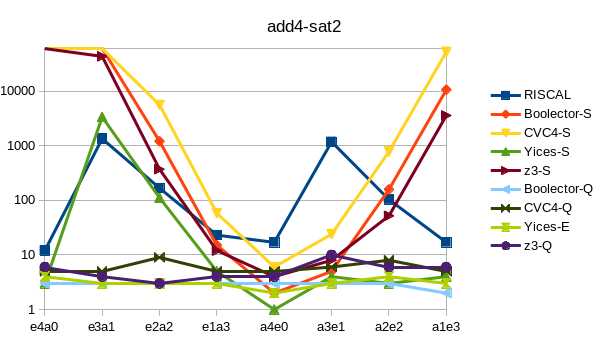} \\
\includegraphics[height=0.155\textheight]{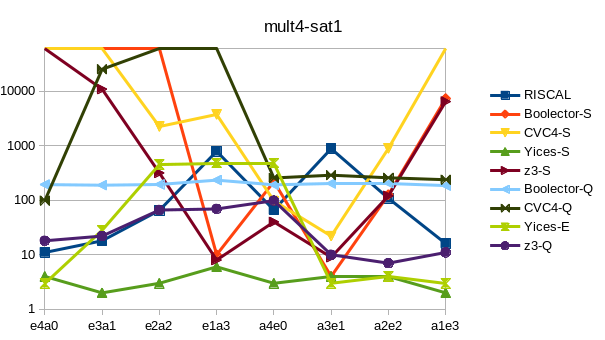} \quad
\includegraphics[height=0.155\textheight]{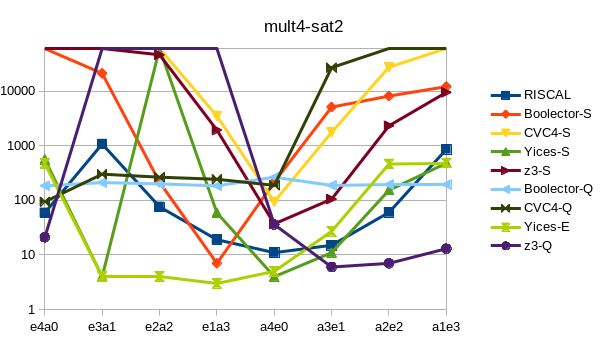} \\
\includegraphics[height=0.155\textheight]{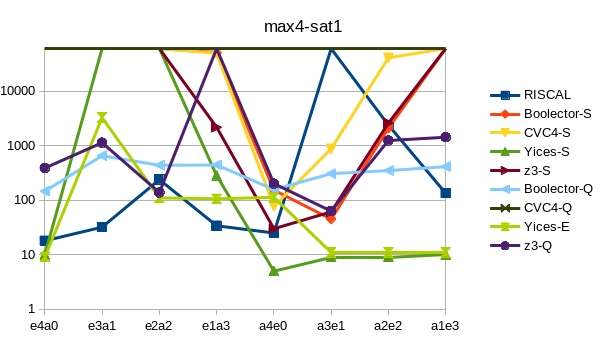} \quad
\includegraphics[height=0.155\textheight]{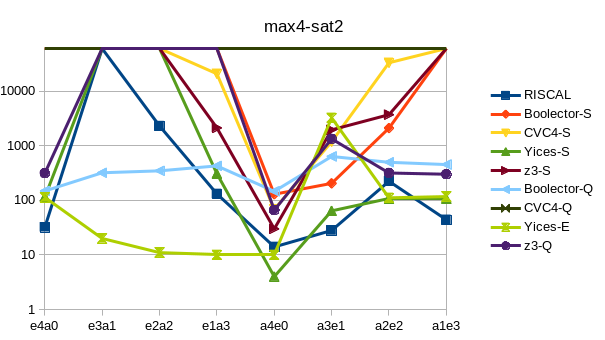} \\
\includegraphics[height=0.155\textheight]{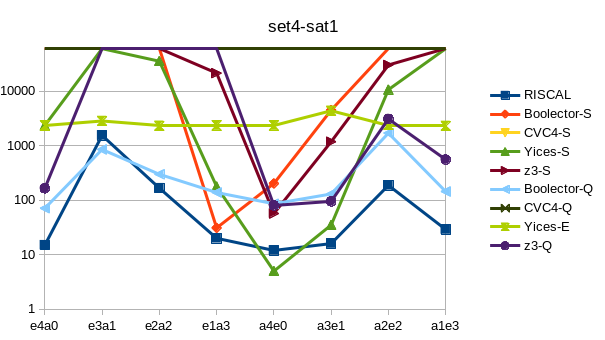} \quad
\includegraphics[height=0.155\textheight]{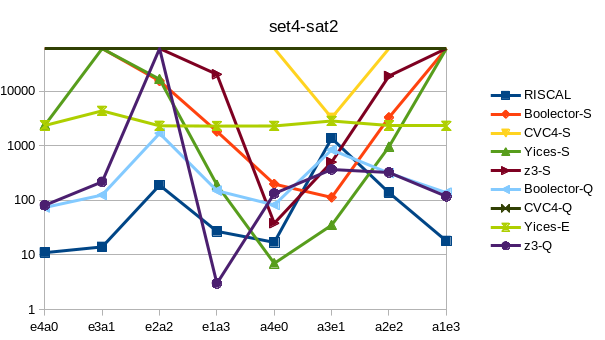} \\
\includegraphics[height=0.155\textheight]{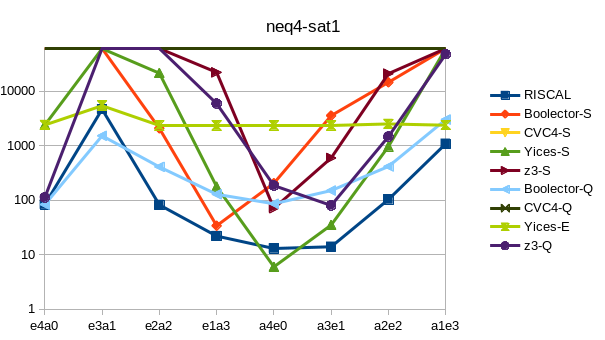} \quad
\includegraphics[height=0.155\textheight]{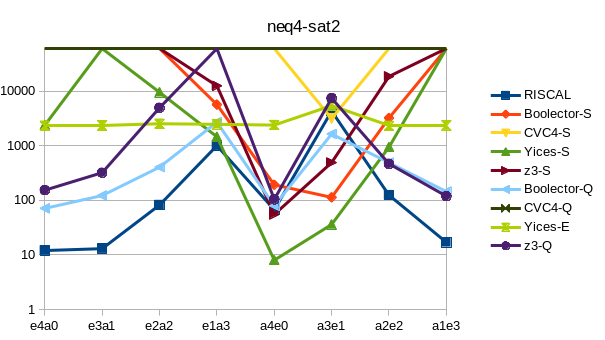} \\
\includegraphics[height=0.155\textheight]{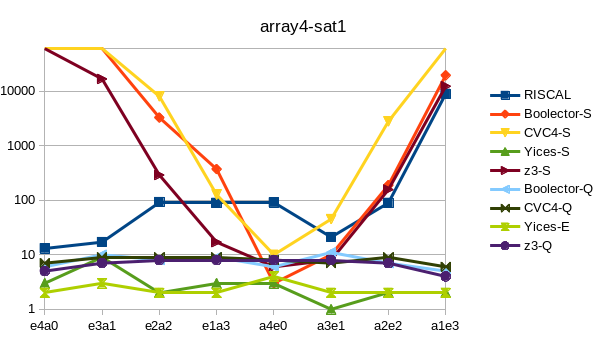} \quad
\includegraphics[height=0.155\textheight]{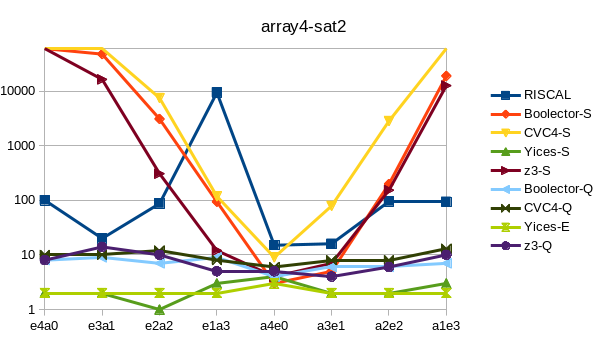} 
\end{center}
\caption{Artificial Benchmarks: Satisfiable Predicates and their Negations}
\label{fig:sat}
\end{figure}

\begin{figure}[H]
\begin{center}
\includegraphics[width=0.49\textwidth]{pictures/exists1.png} 
\includegraphics[width=0.49\textwidth]{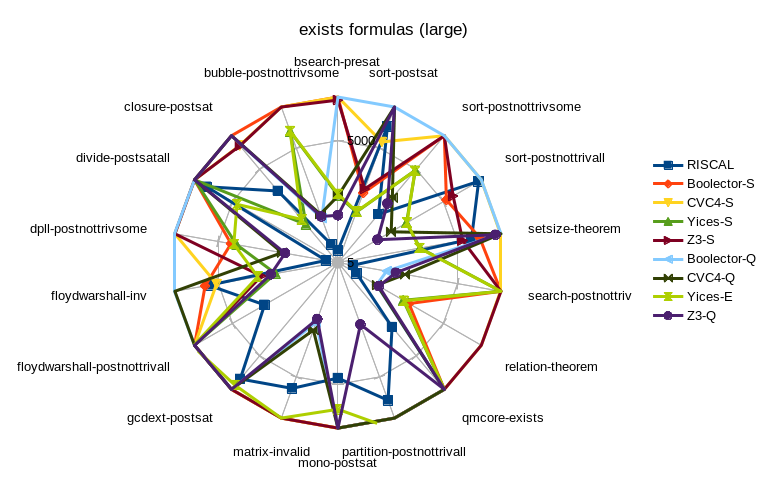} \\ 
\includegraphics[width=0.49\textwidth]{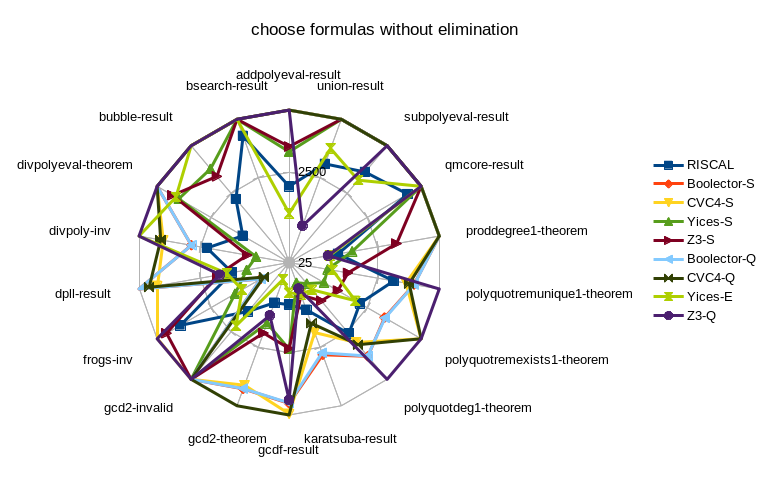} 
\includegraphics[width=0.49\textwidth]{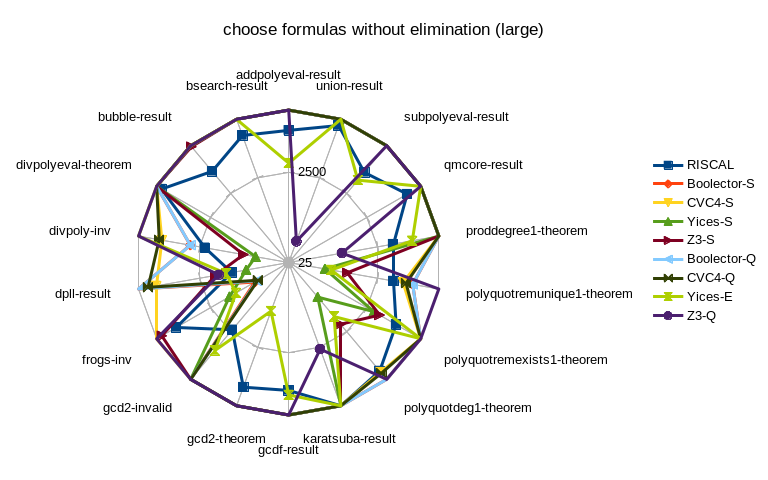} \\
\includegraphics[width=0.49\textwidth]{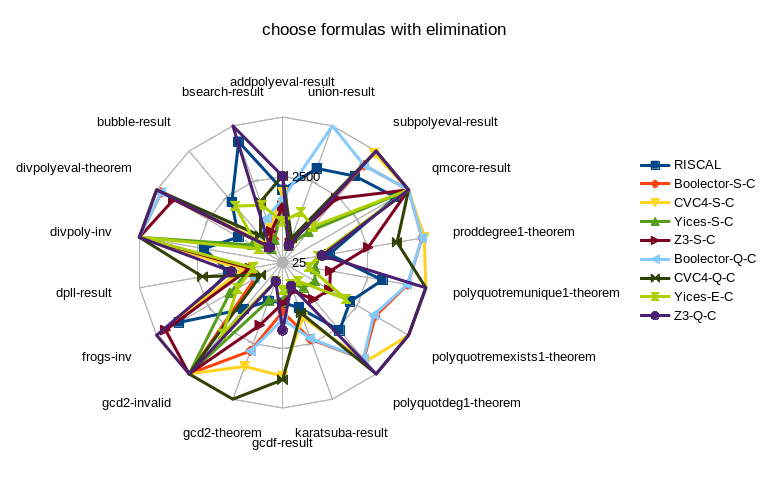} 
\includegraphics[width=0.49\textwidth]{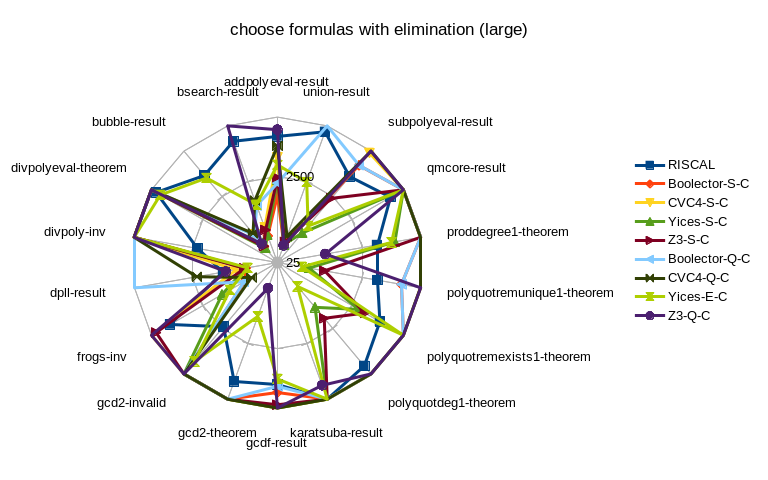} \\
\includegraphics[width=0.49\textwidth]{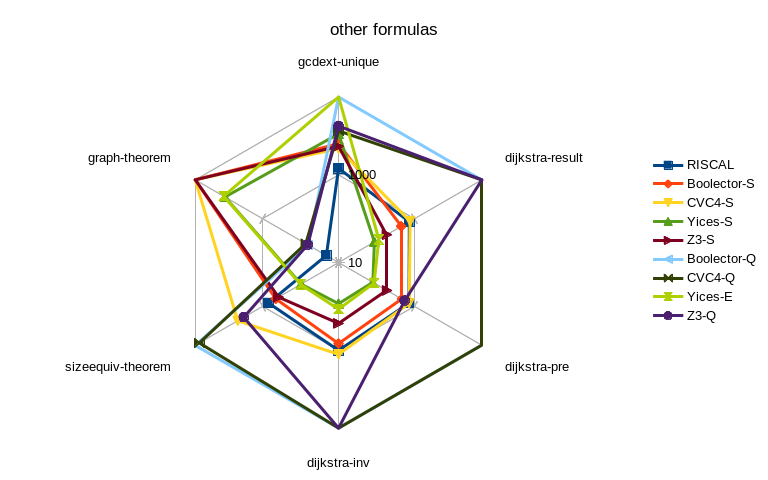} 
\includegraphics[width=0.49\textwidth]{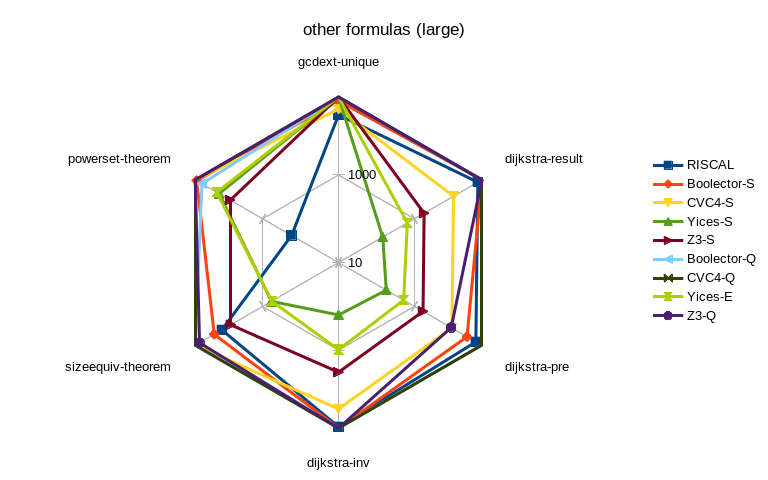} 
\end{center}
\caption{Real-Life Benchmarks (available from 
	\url{https://www.risc.jku.at/research/formal/software/RISCAL/papers/EvalSMT2021-models.tgz})}
\label{fig:real2}
\end{figure}

%% file: main.bbl
\begin{thebibliography}{10}
\providecommand{\bibitemdeclare}[2]{}
\providecommand{\surnamestart}{}
\providecommand{\surnameend}{}
\providecommand{\urlprefix}{Available at }
\providecommand{\url}[1]{\texttt{#1}}
\providecommand{\href}[2]{\texttt{#2}}
\providecommand{\urlalt}[2]{\href{#1}{#2}}
\providecommand{\doi}[1]{doi:\urlalt{http://dx.doi.org/#1}{#1}}
\providecommand{\bibinfo}[2]{#2}

\bibitemdeclare{book}{Ahrendt2016}
\bibitem{Ahrendt2016}
\bibinfo{editor}{Wolfgang \surnamestart Ahrendt\surnameend},
  \bibinfo{editor}{Bernhard \surnamestart Beckert\surnameend},
  \bibinfo{editor}{Richard \surnamestart Bubel\surnameend},
  \bibinfo{editor}{Reinher \surnamestart Hähnle\surnameend},
  \bibinfo{editor}{Peter~H. \surnamestart Schmitt\surnameend} \&
  \bibinfo{editor}{Mattias \surnamestart Ulbrich\surnameend}, editors
  (\bibinfo{year}{2016}): \emph{\bibinfo{title}{{Deductive Software
  Verification --- The KeY Book --- From Theory to Practice}}}.
\newblock {\sl \bibinfo{series}{LNCS}} \bibinfo{volume}{10001},
  \bibinfo{publisher}{Springer International Publishing},
  \bibinfo{address}{Cham}, \doi{10.1007/978-3-319-49812-6}.

\bibitemdeclare{inproceedings}{Barnett2005b}
\bibitem{Barnett2005b}
\bibinfo{author}{Mike \surnamestart Barnett\surnameend},
  \bibinfo{author}{Bor-Yuh~Evan \surnamestart Chang\surnameend},
  \bibinfo{author}{Robert \surnamestart DeLine\surnameend},
  \bibinfo{author}{Bart \surnamestart Jacobs\surnameend} \&
  \bibinfo{author}{K.~Rustan~M. \surnamestart Leino\surnameend}
  (\bibinfo{year}{2005}): \emph{\bibinfo{title}{{Boogie: A Modular Reusable
  Verifier for Object-Oriented Programs}}}.
\newblock In: {\sl \bibinfo{booktitle}{{FMCO 2005: Formal Methods for
  Components and Objects}}}, {\sl \bibinfo{series}{LNCS}}
  \bibinfo{volume}{4111}, \bibinfo{publisher}{Springer},
  \bibinfo{address}{Berlin, Germany}, pp. \bibinfo{pages}{364--387},
  \doi{10.1007/11804192_17}.

\bibitemdeclare{misc}{SMTLIB}
\bibitem{SMTLIB}
\bibinfo{author}{Clark \surnamestart Barrett\surnameend},
  \bibinfo{author}{Pascal \surnamestart Fontaine\surnameend} \&
  \bibinfo{author}{Cesare \surnamestart Tinelli\surnameend}
  (\bibinfo{year}{2016}): \emph{\bibinfo{title}{{The Satisfiability Modulo
  Theories Library (SMT-LIB)}}}.
\newblock \urlprefix\url{http://www.SMT-LIB.org}.

\bibitemdeclare{inproceedings}{Blanchette2011}
\bibitem{Blanchette2011}
\bibinfo{author}{Jasmin~Christian \surnamestart Blanchette\surnameend},
  \bibinfo{author}{Sascha \surnamestart Böhme\surnameend} \&
  \bibinfo{author}{Lawrence~C. \surnamestart Paulson\surnameend}
  (\bibinfo{year}{2011}): \emph{\bibinfo{title}{{Extending Sledgehammer with
  SMT Solvers}}}.
\newblock In: {\sl \bibinfo{booktitle}{CADE-23: Automated Deduction, 23rd
  International Conference}}, {\sl \bibinfo{series}{LNCS}}
  \bibinfo{volume}{6803}, \bibinfo{publisher}{Springer},
  \bibinfo{address}{Berlin, Germany}, pp. \bibinfo{pages}{116--130},
  \doi{10.1007/978-3-642-22438-6_11}.

\bibitemdeclare{inproceedings}{Blanchette2010}
\bibitem{Blanchette2010}
\bibinfo{author}{Jasmin~Christian \surnamestart Blanchette\surnameend} \&
  \bibinfo{author}{Tobias \surnamestart Nipkow\surnameend}
  (\bibinfo{year}{2010}): \emph{\bibinfo{title}{{Nitpick: A Counterexample
  Generator for Higher-Order Logic Based on a Relational Model Finder}}}.
\newblock In: {\sl \bibinfo{booktitle}{ITP 2010: Interactive Theorem Proving}},
  {\sl \bibinfo{series}{LNCS}} \bibinfo{volume}{6172},
  \bibinfo{publisher}{Springer}, \bibinfo{address}{Berlin, Germany}, pp.
  \bibinfo{pages}{131--146}, \doi{10.1007/978-3-642-14052-5_11}.

\bibitemdeclare{inproceedings}{Deharbe2012}
\bibitem{Deharbe2012}
\bibinfo{author}{David \surnamestart Déharbe\surnameend},
  \bibinfo{author}{Pascal \surnamestart Fontaine\surnameend},
  \bibinfo{author}{Yoann \surnamestart Guyot\surnameend} \&
  \bibinfo{author}{Laurent \surnamestart Voisin\surnameend}
  (\bibinfo{year}{2012}): \emph{\bibinfo{title}{{SMT Solvers for Rodin}}}.
\newblock In: {\sl \bibinfo{booktitle}{ABZ 2012: Abstract State Machines,
  Alloy, B, VDM, and Z, Third International Conference, Pisa, Italy, June
  18--21, 2012}}, {\sl \bibinfo{series}{LNCS}} \bibinfo{volume}{7316},
  \bibinfo{publisher}{Springer}, \bibinfo{address}{Berlin, Germany}, pp.
  \bibinfo{pages}{194--207}, \doi{10.1007/978-3-642-30885-7_14}.

\bibitemdeclare{inproceedings}{Ekici2017}
\bibitem{Ekici2017}
\bibinfo{author}{Burak \surnamestart Ekici\surnameend}, \bibinfo{author}{Alain
  \surnamestart Mebsout\surnameend}, \bibinfo{author}{Cesare \surnamestart
  Tinelli\surnameend}, \bibinfo{author}{Chantal \surnamestart
  Keller\surnameend}, \bibinfo{author}{Guy \surnamestart Katz\surnameend}
  et~al. (\bibinfo{year}{2017}): \emph{\bibinfo{title}{{SMTCoq: A Plug-In for
  Integrating SMT Solvers into Coq}}}.
\newblock In: {\sl \bibinfo{booktitle}{CAV 2017: Computer Aided Verification,
  Heidelberg, Germany, July 24--28, 2017}}, {\sl \bibinfo{series}{LNCS}}
  \bibinfo{volume}{10427}, \bibinfo{publisher}{Springer},
  \bibinfo{address}{Cham, Switzerland}, pp. \bibinfo{pages}{126--133},
  \doi{10.1007/978-3-319-63390-9_7}.

\bibitemdeclare{misc}{Ghazi2015}
\bibitem{Ghazi2015}
\bibinfo{author}{Aboubakr Achraf~El \surnamestart Ghazi\surnameend} \&
  \bibinfo{author}{Mana \surnamestart Taghdiri\surnameend}
  (\bibinfo{year}{2015}): \emph{\bibinfo{title}{{Analyzing Alloy Formulas using
  an SMT Solver: A Case Study}}}.
\newblock \bibinfo{howpublished}{AFM10: Automated Formal Methods, July 14,
  2010, Edinburgh, UK}.
\newblock \urlprefix\url{https://arxiv.org/abs/1505.00672}.

\bibitemdeclare{inproceedings}{Jackson2000}
\bibitem{Jackson2000}
\bibinfo{author}{Daniel \surnamestart Jackson\surnameend}
  (\bibinfo{year}{2000}): \emph{\bibinfo{title}{{Automating First-Order
  Relational Logic}}}.
\newblock In: {\sl \bibinfo{booktitle}{SIGSOFT'00/FSE-8 International
  Symposium, San Diego, California, USA, November 2000}}, {\sl
  \bibinfo{series}{SIGSOFT Software Engineering Notes}}
  \bibinfo{volume}{25(6)}, \bibinfo{publisher}{ACM}, \bibinfo{address}{New
  York, NY, USA}, pp. \bibinfo{pages}{130--139}, \doi{10.1145/355045.355063}.

\bibitemdeclare{inproceedings}{Leino2016}
\bibitem{Leino2016}
\bibinfo{author}{K.~Rustan~M. \surnamestart Leino\surnameend}
  (\bibinfo{year}{2010}): \emph{\bibinfo{title}{{Dafny: An Automatic Program
  Verifier for Functional Correctness}}}.
\newblock In \bibinfo{editor}{Edmund~M. \surnamestart Clarke\surnameend} \&
  \bibinfo{editor}{Andrei \surnamestart Voronkov\surnameend}, editors: {\sl
  \bibinfo{booktitle}{LPAR-16: Logic Programming and Automated Reasoning, 16th
  International Conference, Dakar, Senegal, April 25--May 1, 2010}}, {\sl
  \bibinfo{series}{LNCS}} \bibinfo{volume}{6355}, \bibinfo{publisher}{Springer,
  Berlin, Germany}, pp. \bibinfo{pages}{{348--370}},
  \doi{10.1007/978-3-642-17511-4_20}.

\bibitemdeclare{inproceedings}{Lin2017}
\bibitem{Lin2017}
\bibinfo{author}{Hsin-Hung \surnamestart Lin\surnameend} \&
  \bibinfo{author}{Bow-Yaw \surnamestart Wang\surnameend}
  (\bibinfo{year}{2017}): \emph{\bibinfo{title}{{Releasing VDM Proof
  Obligations with SMT Solvers}}}.
\newblock In: {\sl \bibinfo{booktitle}{MEMOCODE '17: 15th ACM-IEEE
  International Conference on Formal Methods and Models for System Design,
  Vienna, Austria, September 29--October 2, 2017}}, \bibinfo{publisher}{ACM},
  \bibinfo{address}{New York, NY, USA}, p. \bibinfo{pages}{132–135},
  \doi{10.1145/3127041.3127066}.

\bibitemdeclare{inproceedings}{Merz2016}
\bibitem{Merz2016}
\bibinfo{author}{Stephan \surnamestart Merz\surnameend} \&
  \bibinfo{author}{Hernán \surnamestart Vanzetto\surnameend}
  (\bibinfo{year}{2016}): \emph{\bibinfo{title}{{Encoding TLA$^{+}$ into
  Many-Sorted First-Order Logic}}}.
\newblock In: {\sl \bibinfo{booktitle}{ABZ 2016: Abstract State Machines,
  Alloy, B, TLA, VDM, and Z: 5th International Conference, Linz, Austria, May
  23--27, 2016}}, {\sl \bibinfo{series}{LNCS}} \bibinfo{volume}{9675},
  \bibinfo{publisher}{Springer}, \bibinfo{address}{Cham, Switzerland}, pp.
  \bibinfo{pages}{54--69}, \doi{10.1007/978-3-319-33600-8_3}.

\bibitemdeclare{inproceedings}{Reger2017}
\bibitem{Reger2017}
\bibinfo{author}{Giles \surnamestart Reger\surnameend}, \bibinfo{author}{Martin
  \surnamestart Suda\surnameend} \& \bibinfo{author}{Andrei \surnamestart
  Voronkov\surnameend} (\bibinfo{year}{2017}):
  \emph{\bibinfo{title}{{Instantiation and Pretending to be an SMT Solver with
  Vampire}}}.
\newblock In: {\sl \bibinfo{booktitle}{SMT 2017 Workshop, Heidelberg, Germany,
  July 22--23, 2017}}, {\sl \bibinfo{series}{CEUR Workshop Proceedings}}
  \bibinfo{volume}{1889}, pp. \bibinfo{pages}{63--75}.
\newblock \urlprefix\url{http://ceur-ws.org/Vol-1889/paper6.pdf}.

\bibitemdeclare{mastersthesis}{Reichl2020}
\bibitem{Reichl2020}
\bibinfo{author}{Franz-Xaver \surnamestart Reichl\surnameend}
  (\bibinfo{year}{2020}): \emph{\bibinfo{title}{{The Integration of SMT Solvers
  into the RISCAL Model Checker}}}.
\newblock Master's thesis, \bibinfo{school}{Research Institute for Symbolic
  Computation (RISC)}, \bibinfo{address}{Johannes Kepler University Linz,
  Austria}.
\newblock
  \urlprefix\url{https://www.risc.jku.at/publications/download/risc_6103/Thesis.pdf}.

\bibitemdeclare{inproceedings}{Schreiner2018b}
\bibitem{Schreiner2018b}
\bibinfo{author}{Wolfgang \surnamestart Schreiner\surnameend}
  (\bibinfo{year}{2018}): \emph{\bibinfo{title}{{Validating Mathematical
  Theories and Algorithms with RISCAL}}}.
\newblock In: {\sl \bibinfo{booktitle}{{CICM 2018, 11th Conference on
  Intelligent Computer Mathematics, Hagenberg, Austria, August 13--17}}}, {\sl
  \bibinfo{series}{LNCS/Lecture Notes in Artificial Intelligence}}
  \bibinfo{volume}{11006}, \bibinfo{publisher}{Springer, Berlin}, pp.
  \bibinfo{pages}{248--254}, \doi{10.1007/978-3-319-96812-4_21}.

\bibitemdeclare{misc}{RISCAL}
\bibitem{RISCAL}
\bibinfo{author}{Wolfgang \surnamestart Schreiner\surnameend}
  (\bibinfo{year}{2019}): \emph{\bibinfo{title}{{The RISC Algorithm Language
  (RISCAL)}}}.
\newblock \bibinfo{howpublished}{Research Institute for Symbolic Computation
  (RISC), Johannes Kepler University, Linz, Austria}.
\newblock
  \urlprefix\url{https://www.risc.jku.at/research/formal/software/RISCAL}.

\bibitemdeclare{article}{Schreiner2020c}
\bibitem{Schreiner2020c}
\bibinfo{author}{Wolfgang \surnamestart Schreiner\surnameend} \&
  \bibinfo{author}{Franz-Xaver \surnamestart Reichl\surnameend}
  (\bibinfo{year}{2020}): \emph{\bibinfo{title}{{Mathematical Model Checking
  Based on Semantics and SMT}}}.
\newblock {\sl \bibinfo{journal}{Transactions on Internet Research}}
  \bibinfo{volume}{16}(\bibinfo{number}{2}), pp. \bibinfo{pages}{4--13}.
\newblock
  \urlprefix\url{http://ipsitransactions.org/journals/papers/tir/2020jul/p2.pdf}.

\bibitemdeclare{techreport}{Schreiner2021}
\bibitem{Schreiner2021}
\bibinfo{author}{Wolfgang \surnamestart Schreiner\surnameend} \&
  \bibinfo{author}{Franz-Xaver \surnamestart Reichl\surnameend}
  (\bibinfo{year}{2021}): \emph{\bibinfo{title}{{Semantic Evaluation versus SMT
  Solving in the RISCAL Model Checker}}}.
\newblock \bibinfo{type}{Technical Report} \bibinfo{number}{21-11},
  \bibinfo{institution}{RISC}, \bibinfo{address}{Johannes Kepler University,
  Linz, Austria}.
\newblock
  \urlprefix\url{https://www.risc.jku.at/publications/download/risc_6328/21-11.pdf}.

\bibitemdeclare{article}{Schreiner2020b}
\bibitem{Schreiner2020b}
\bibinfo{author}{Wolfgang \surnamestart Schreiner\surnameend},
  \bibinfo{author}{William \surnamestart Steingartner\surnameend} \&
  \bibinfo{author}{Valerie \surnamestart Novitzká\surnameend}
  (\bibinfo{year}{2020}): \emph{\bibinfo{title}{{A Novel Categorical Approach
  to the Semantics of Relational First-Order Logic}}}.
\newblock {\sl \bibinfo{journal}{Symmetry}}
  \bibinfo{volume}{12}(\bibinfo{number}{10}), \doi{10.3390/sym12101584}.

\bibitemdeclare{misc}{SMT-COMP}
\bibitem{SMT-COMP}
\bibinfo{author}{\surnamestart SMT-COMP\surnameend} (\bibinfo{year}{2021}):
  \emph{\bibinfo{title}{{SMT-COMP: The International Satisfiability Modulo
  Theories (SMT) Competition.}}}
\newblock \urlprefix\url{https://smt-comp.github.io}.

\bibitemdeclare{inproceedings}{Torlak2007}
\bibitem{Torlak2007}
\bibinfo{author}{Emina \surnamestart Torlak\surnameend} \&
  \bibinfo{author}{Daniel \surnamestart Jackson\surnameend}
  (\bibinfo{year}{2007}): \emph{\bibinfo{title}{{Kodkod: A Relational Model
  Finder}}}.
\newblock In: {\sl \bibinfo{booktitle}{TACAS 2007: Tools and Algorithms for the
  Construction and Analysis of Systems, 3th International Conference, Braga,
  Portugal, March 24--April 1, 2007}}, {\sl \bibinfo{series}{LNCS}}
  \bibinfo{volume}{4424}, \bibinfo{publisher}{Springer},
  \bibinfo{address}{Berlin, Germany}, pp. \bibinfo{pages}{632--647},
  \doi{10.1007/978-3-540-71209-1_49}.

\end{thebibliography}
